\documentclass[twocolumn,prl,aps,showpacs]{revtex4}
\usepackage{graphicx,graphics}
\usepackage{enumerate}
\usepackage{subfigure}
\usepackage{amsmath}
\usepackage{amsfonts}
\usepackage{amssymb}
\usepackage{amsthm}
\usepackage{psfrag}

\def\D{\Delta}

\def\R{{\cal R}}

\def\Z{\mathbb{Z}}
\def\la{\langle}
\def\ra{\rangle}
\renewcommand\L{\mathcal{L}}
\def\C{\mathcal C}

\def\be{\begin{equation}}
\def\ee{\end{equation}}
\def\bea{\begin{eqnarray}}
\def\eea{\end{eqnarray}}

\begin{document}

\title{Pair correlations in sandpile model: a check of logarithmic conformal field theory}

\author{V.S. Poghosyan$^1$, S.Y. Grigorev$^1$, V.B. Priezzhev$^1$ and P. Ruelle$^2$}
\affiliation{
$^1$Bogoliubov Laboratory of Theoretical Physics, JINR, 141980 Dubna, Russia\\
$^2$Institut de Physique Th\'{e}orique, Universit\'{e} catholique de Louvain, B-1348
Louvain-La-Neuve, Belgium}

\begin{abstract}
We compute the correlations of two height variables in the two-dimensional
Abelian sandpile model. We extend the known result for two minimal heights
to the case when one of the heights is bigger than one.
We find that the most dominant correlation $\log r/r^4$
exactly fits the prediction obtained within the logarithmic conformal approach.
%
\end{abstract}


\maketitle

Conformal field theory has proved to be extraordinarily powerful in the description of
universality classes of equilibrium critical models
in two dimensions \cite{cft}. Critical exponents, correlation functions,
finite-size scaling, perturbations and boundary conditions, among others, have all been
studied within the conformal approach, and thoroughly (and successfully) compared with
numerical data.

More recently, increased interest has been turned toward logarithmic conformal theories, as a
larger class of conformal theories, interesting in its own right, but also as a description of
certain non-equilibrium lattice models. In particular, dense polymers \cite{polym,log}, sandpile
models \cite{sand,jpr} and percolation \cite{perc,log} are lattice realizations of logarithmic
conformal theories. An infinite
series
of such lattice models have been defined in \cite{log}.

The logarithmic theories are however much lesser understood than the more usual, non-logarithmic
ones. This is due to their higher level of complexity, which somehow reflect the complexity of
the associated lattice models. Indeed the models mentioned above all have intrinsic non-local
features. In this respect, it may appear to be strange, if not miraculous, that a lattice model
with non-local variables can be described, in the scaling limit, by a local field theory. The
only trace the lattice non-localities leave in the continuum local theory seems to be the
presence of logarithms in correlation functions.

It is therefore essential to check that the logarithmic conformal description is indeed
appropriate for these models, as extensively as it has been done for equilibrium critical
phenomena (see for instance \cite{henkel}).

It is our purpose in this Letter to take further steps in this necessary procedure, in the
context of the two-dimensional Abelian sandpile model. A certain number of checks have been
carried out for this model (see \cite{sand}), but the one we propose here is more crucial
because it deals with microscopic variables which are manisfestly
non-local, and for which the logarithmic conformal theory makes a very definite prediction.
It therefore exposes in the clearest possible way the non-local features of the model.

Namely, we compute, in the infinite discrete plane, the 2-site correlations
$P_{1i}(r)-P_1P_i$ of two height variables, one of which being equal to 1, the other,
$h_i$, being equal to 2, 3 or 4 (here $P_i$ is the 1-site probability on the infinite plane).
Conformal field theory predicts that the dominant term of these is given by \cite{jpr}
\be
P_{1i}(r) - P_1 P_i = c_i {\log r \over r^4} + \ldots,
\ee
with known coefficients $c_i$. New and explicit lattice calculations,
to be detailed below, fully confirm these results, and exactly reproduces the coefficients $c_i$.

Logarithmic conformal theory also predicts that the 2-site correlations $P_{ij}(r)-P_iP_j$
of two heights bigger or equal to 2 decay like ${\log^2 r/r^4}$, but the explicit lattice
calculation of these remains out of range for the moment.


\section{The sandpile model and logarithmic conformal theory}
We briefly recall the sandpile model introduced by Bak, Tang and Wiesenfeld in \cite{btw}
(see \cite{ipd} for further details).

Every site $i$ of a finite rectangular grid $\L$ is assigned a height variable $h_i$,
taking the four values $1, 2, 3$ and 4. A configuration $\C$ is the set of values
 $\{h_i\}$ for all sites. A discrete stochastic dynamics is defined on the set of
configurations. If $\C_t$ is the configuration at time $t$, the height at a random
site $i$ of $\C_t$ is incremented by 1, $h_i \to h_i + 1$, making a new configuration
 $\C'_t$. If the (new) height $h_i$ in $\C'_t$ is smaller or equal to 4, one simply
sets $\C_{t+1}=\C'_t$. If not, all sites $j$ such that their height variables $h_j$
exceed 4 topple, a process by which $h_j$ is decreased by 4, and the height of all
the nearest neighbours of $j$ are increased by 1. That is, when the site $j$ topples,
the heights are updated according to
\be
h_i \to h_i - \Delta_{ji},
\ee
with $\Delta$ the discrete Laplacian, $\Delta_{ii}=4$, $\Delta_{ij}=-1$ for nearest
neighbour sites, and $\Delta_{ij}=0$ otherwise. This toppling process stops when all
height variables are between 1 and 4; the configuration so obtained defines $\C_{t+1}$.

The boundary sites are dissipative, because a toppling there evacuates
one or two grains of sand, which we imagine are collected in a sink site, connected to
all dissipative sites. The presence of dissipative sites is essential for the dynamics
to be well-defined, since it makes sure that the toppling process stops in a finite time.

When the dynamics is run over long periods, the sandpile builds up,
being
subjected to avalanches spanning large portions of the system. This correlates the height
variables over very large distances, and makes the system critical in the thermodynamic limit.

It turns out that, when the dynamics is run for long enough, and no matter what the initial
configuration is, the
sandpile
enters a stationary regime, in which only special configurations
occur with equal probability, the so-called recurrent configurations \cite{ddhar}.
The recurrent set $\R$ forms a small fraction of all configurations, since
\be
|\R| = \det\D \simeq (3.21)^N,
\ee
where $N$ is the number of sites. So the asymptotic state of the sandpile is controlled by
a unique invariant distribution $P_{\L}^*$, uniform on the set $\R$ of recurrent configurations,
and zero on the non-recurrent (transient) ones. In the infinite volume limit, the invariant
measure $P^*_\L$ is believed to become a conformal field theoretic measure.

To be recurrent, the height values of a configuration must satisfy certain global conditions
\cite{ddhar}, leading to non-local features. For what follows, it will be enough to know that
the recurrent configurations are in one-to-one correspondence with oriented spanning trees
on $\L^\star$, the original lattice $\L$ supplemented with the sink site. This change of
variables, more convenient to perform actual calculations, also yields a different lighting
on the non-localities of the model.

Spanning trees are acyclic configurations of arrows: at each site $i$ of $\L$, there is an
outgoing arrow, pointing to any one of its $\D_{ii}$ neighbours (if $i$ is dissipative, the
arrow can point to the sink site). A configuration of arrows defines a spanning tree if it
contains no loop. By construction, the paths formed by the arrows all lead to the sink site
$\star$, which is the root of the tree.

The mapping between recurrent configurations and trees is complicated and non-local; however
the spanning trees provide an equivalent description. The global conditions that the heights
of recurrent configurations have to satisfy are encoded in the property of arrow configurations
of containing no loop, also a global constraint. The invariant measure $P^*_\L$ becomes simply
a uniform distribution on the spanning trees.

Height values at a given site can be related to properties of spanning trees. To do so, one
defines the notion of predecessor: a site $j$ is a predecessor of $i$ if the unique path
from $j$ to the root passes through $i$. Then it has been shown \cite{Priez} that the trees
in which the site $i$ (not on the boundary) has exactly $a-1$ predecessors among its nearest
neighbours correspond to configurations where $h_i \geq a$, for $a=1,2,3$ or 4. So
configurations with $h_i=1$ are associated with trees which have a leaf at $i$; this is a
local property which may be verified by looking at the neighbourhood of $i$ only. In contrast,
heights 2, 3 and 4 correspond to non-local properties in terms of the trees.

Using this correspondence, joint probabilities for heights $P[h_i=a,h_j=b,\ldots]$ can be
related to the fractions of trees satisfying certain conditions regarding the number of
predecessors of $i,j,\ldots$ among their nearest neighbours. However, because of the
remark we have just made, probabilities with heights 1 only are considerably easier than
those involving higher heights. So far, the only probabilities involving higher heights in
the bulk which have been computed are the 1-site probabilities $P[h_i=a]$ on the upper-half
plane \cite{jpr}. They provided enough input to assess the conformal nature of the four height
variables in the scaling limit.

The logarithmic conformal theory, relevant to the sandpile model, has central charge $c=-2$.
Among the distinctive features of a logarithmic theory is the presence of reducible yet
indecomposable Virasoro representations; this property in turn introduces logarithms in their
correlators \cite{lcft}.

The fields describing the scaling limit of the four lattice height variables
$\delta(h_z-i)-P_i$, which we call $h_i(z)$, have been determined in \cite{jpr}.
As hinted by the remarks made above, the height 1 field is very different from the
other heights' fields. It turns out that $h_1$ is a primary field with conformal
weights $(1,1)$, while the other three, $h_2,h_3$ and $h_4$, are all related to a
single field, identified with the logarithmic partner of $h_1$. More precisely, if $h_1$ is
the primary field normalized as the height 1 variable on the lattice, then $h_2$ satisfies
the triangular relations,
\be
L_0 h_2 = h_2 - {1 \over 2} h_1, \quad L_1 h_2 = \rho, \quad L_{-1} \rho = -{1 \over 4}h_1,
\ee
where $\rho$ is a (0,1) field. In fact, $h_1$ and $h_2$ are members of the non-chiral
version of the indecomposable representation called $\R_{2,1}$ in \cite{gk2}. The last
two fields are linear combinations,
$h_3 = \alpha_3 h_2 + \beta_3 h_1$, $h_4 = \alpha_4 h_2 + \beta_4 h_1$,
and may also be viewed as logarithmic partners of $h_1$. The coefficients
$\alpha_i,\beta_i$ are such that $h_3$ and $h_4$, like $h_1$ and $h_2$, have the
same normalization as their lattice counterparts; their exact values are known \cite{jpr}.

The identification of the height fields makes it possible to compute correlations.
In particular the joint probabilities for two height variables on the infinite plane
$\Z^2$ correspond, in the scaling limit, to 3-point correlators in the conformal theory,
\be
P_{ij}(z_1,z_2) - P_iP_j = \la h_i(z_1) h_j(z_2) \omega(\infty) \ra,
\ee
where $\omega$ is a weight (0,0) conformal field, logarithmic partner of the identity
\cite{jpr}. Indeed the infinite plane should be thought of as the limit of a growing
finite grid, which has dissipation located along the boundary. In the infinite volume
limit, the boundaries, and with them, the dissipation, are sent off to infinity. The
field $\omega$ precisely realizes the insertion of dissipation at infinity, required for the
sandpile model to be well-defined.

The 3-point correlators have been computed in \cite{jpr}, and take the general form
($z_{12} \equiv z_1-z_2$)
\be
\la h_i(z_1) h_j(z_2) \omega(\infty) \ra = {A_{ij} + B_{ij} \log|z_{12}| + C_{ij}
\log^2|z_{12}| \over |z_{12}|^4},
\ee
where $C_{ij}=0$ if $\min(i,j)=1$, and moreover $B_{11}=0$ and $A_{11}=-P_1^2/2$
\cite{Dhar}, so that, depending on $i$ and $j$, one, two or three terms in the numerator
are present. The coefficient of the dominant term, i.e. the largest power
of $\log|z_{12}|$, could be determined exactly, and yields the dominant contribution
of the 2-site probabilities \cite{jpr}
\bea
P_{1i}(r) - P_1P_i &\simeq& -{\alpha_i P_1^2 \over 2r^4} \log r,\qquad i>1, \label{P1i}\\
P_{ij}(r) - P_iP_j &\simeq& -{\alpha_i \alpha_j P_1^2 \over 2r^4} \log^2 r,\qquad i,j>1,
\eea
where $P_1 = 2(\pi-2)/\pi^3$, as first computed in \cite{Dhar}, and
\be
\alpha_2 = 1, \quad \alpha_3 = {8-\pi \over 2(\pi-2)}, \quad \alpha_4 = -{\pi+4 \over 2(\pi-2)}.
\ee


\section{Calculations on the lattice}
\begin{figure}[h!]
\includegraphics[width=60mm]{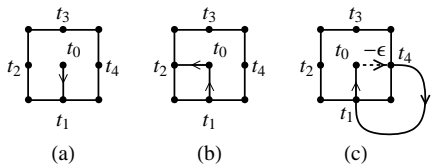}
\caption{\label{fig1} Non-zero elements of $B$ related to $t_0$ (arrowed bonds) for
(a) $\Delta' = \Delta+B_1$, (b) $\Delta'=\Delta_\text{local}$ and
(c) $\Delta'=\Delta_\text{loop}$.
The bond $[t_0,t_4]$ in $\Delta_\text{loop}$ is weighted by $-\varepsilon$.}
\end{figure}

It has been shown in \cite{Dhar} and \cite{Priez} (see also \cite{jpr} for details) that
height probabilities $P_i$ in the ASM can be reduced to the computation of determinants
of discrete Laplacian matrices perturbed by a number of defects. The resulting matrices
$\Delta'=\Delta + B$ differ from the regular Laplacian by a defect matrix $B$, with $B=0$
except for a finite number of elements. Given a lattice point $t_0$, non-zero elements of
$B$ related to $t_0$ can be marked by arrows at adjacent bonds (Fig.\ref{fig1}).
For instance, the non-zero part of the matrix $B=B_1$ used for the evaluation
of $P_1$ (Fig.\ref{fig1}a) is
\begin{equation}
\label{B1}
B_1=
{\small \left(
\begin{array}{ccccc}
-3 & 1  & 1  &  1\\
 1 & -1 & 0  &  0\\
 1 & 0  & -1 &  0\\
 1 & 0  & 0  & -1\\
\end{array}
\right)},
\end{equation}
where rows and columns are labeled by $t_0$, $t_2$, $t_3$, $t_4$. The probability to
have a height $1$ at $t_0$ is then \cite{Dhar}
\begin{equation}
\label{P1}
P_1=\frac{\det(\Delta + B_1)}{\det\Delta}=\det({\mathbb I} + B_1 G),
\end{equation}
where $G=\Delta^{-1}$.
The explicit form of the translation invariant Green function on the plane,
$G(\vec{r})\equiv G_{\vec{r},\vec{0}}=G_{0,0}+g_{p,q}$ for $\vec{r}=(p,q)$,
\begin{equation}
\label{Green}
g_{p,q}=
\frac{1}{8\pi^2}
\int\!\!\!\!\int_{-\pi}^{\pi}
\frac{e^{i p\alpha}e^{i q\beta}-1}{2-\cos\alpha- \cos\beta}{\rm d}\alpha {\rm d}\beta,
\end{equation}
implies $P_1=2(\pi-2)/\pi^3$.

\begin{figure}[h!]
\includegraphics[width=60mm]{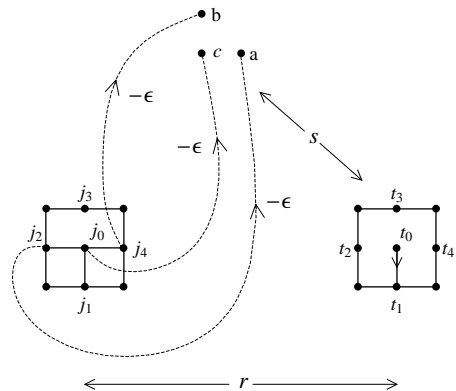}
\caption{\label{fig2} Structure of matrix $\Delta_{1\Theta}$. On the left side, the defects
are the removed bond $[j_3,j_0]$ and three additional bonds
$[j_4,b]$, $[j_0,c]$, $[j_2,a]$ with weight $-\varepsilon$. The right part shows the same
defect as in Fig.\ref{fig1}a.}
\end{figure}

\begin{figure}[h!]
\includegraphics[width=60mm]{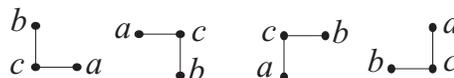}
\caption{\label{fig3} Four possible orientations of the group $[a,b,c]$.}
\end{figure}

The probability to have a height $2$ can be written \cite{Priez} as
\begin{eqnarray}
\label{P2}
P_{2}&=&P_{1}+\frac{4\det\Delta_\text{local}}{\det\Delta}
   + \lim_{\varepsilon\to\infty}\frac{4\det\Delta_\text{loop}}{\varepsilon\det\Delta}+\nonumber\\
  &+&\sum_{[a,b,c]}\lim_{\varepsilon\to\infty}\frac{2\det\Delta_\Theta}{\varepsilon^3\det\Delta},
\end{eqnarray}
where the defect matrices related to $\Delta_\text{local}$ and $\Delta_\text{loop}$ are shown
in Fig.\ref{fig1}b-c, and that related to $\Delta_\Theta$ is on the left side of Fig.\ref{fig2}.
The matrix $\Delta_\Theta$ differs from $\Delta$ by the removed bond $[j_0,j_3]$ and three
additional matrix elements (bonds) weighted by $-\varepsilon$ between the sites $j_0$, $j_2$,
$j_4$ and a triplet of neighbouring sites $[a,b,c]$, whose position and orientation
(see Fig.\ref{fig3}) are to be summed over the whole lattice, with the restriction
that the group $[a,b,c]$ does not overlap with $j_0$, $j_2$, $j_3$, $j_4$.

\begin{figure}[h!]
\includegraphics[width=80mm]{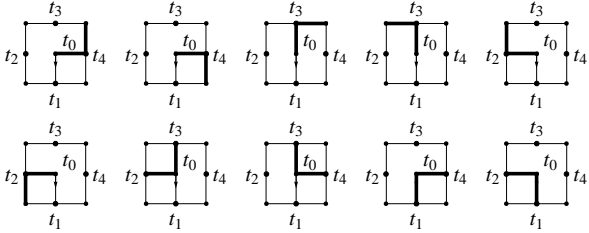}
\caption{\label{fig4} Forbidden positions of the group [a,b,c].}
\end{figure}

The 2-site probability $P_{12}$ combines the defect of $\Delta_1$ with those of
$\Delta_{\text{local}}$, $\Delta_{\text{loop}}$ and $\Delta_{\Theta}$. Simple
calculations show that the matrices $\Delta_{\text{local}}$ and $\Delta_{\text{loop}}$
contribute a term $1/r^4$ to the asymptotics of $P_{12}(r)$ for large $r$, and are
therefore subdominant. Thus the leading contribution comes from the matrix $\Delta_{1\Theta}$
combining the defects of $\Delta_1$ and $\Delta_{\Theta}$, as shown in Fig.\ref{fig2}.
The correlation function $P_{1\Theta}(r)$ is
\begin{equation}
\label{P1Theta}
P_{1\Theta}(r) ={\sum_{[a,b,c]}}'\lim_{\varepsilon\to\infty}\frac{2\det\Delta_{1\Theta}}
{\varepsilon^3\det\Delta},
\end{equation}
where the prime means that the sum excludes the terms where at least one edge in the group
$[a,b,c]$ overlaps a deleted edge adjacent to $t_0$. The ten forbidden positions are shown
in Fig.\ref{fig4}.

The ratios of determinants in Eq.(\ref{P1Theta}) are computed as
$\det({\mathbb I}+B_{1\Theta}G)$, like Eq.(\ref{P1}),
where the non-zero part of $B_{1\Theta} = \Delta_{1\Theta} - \Delta$ is a $8 \times 9$
block diagonal matrix. The first block is
\begin{equation}
B_\Theta=
{\small \left(
\begin{array}{ccccc}
 1 & -1&       0      &        0      &      0     \\
 0 & 0 & -\varepsilon &        0      &      0     \\
 0 & 0 &       0      &  -\varepsilon &      0     \\
 0 & 0 &       0      &        0      &-\varepsilon\\
\end{array}
\right)},
\end{equation}
with rows $j_3,j_0,j_2,j_4$ and columns $j_0,j_3,c,a,b$, while the second block is $B_1$ in
Eq.(\ref{B1}). For large $r \gg 1$, we can replace all Green functions containing $r$ by their
asymptotic value,

\begin{equation}
g_{p,q}=-\frac{\ln(p^2+q^2)}{4\pi} - \frac{1}{\pi} \left( \frac{\gamma}{2} + \frac{3}{4} \log 2 \right),
\end{equation}
where $p^2+q^2 \gg 1$ and $\gamma=0.57721\ldots$ is the Euler constant.
Expanding the determinants in Eq.(\ref{P1Theta}), we obtain, for the sum over the forbidden
positions,
\begin{equation}
F(r) = \frac{2(\pi-2)^2}{\pi^6}\frac{\log r}{r^4} + O\left(\frac{1}{r^4}\right).
\end{equation}
We can now write the desired correlation in the form
\begin{equation}
P_{1\Theta}(r) - P_{1} P_{\Theta} = 2 \left( \sum_{\vec{s}} \mathcal{U}_r(\vec{s}) - F(r)\right),
\end{equation}
where the sum is taken over all lattice points $\vec{s}=(k,l)$ and $P_{\Theta}$ is the last
term in Eq.(\ref{P2}). The function $\mathcal{U}_r(\vec{s})$ behaves as
$r^{-4}$, if $s > r \gg 1$ and as $s^{-4}r^{-4}\log r$, if $r \gg 1$, $s \gg 1$, $s < r$.
In the region $r \gg 1$ and $s < r$, we have
\begin{equation}
\mathcal{U}_r(\vec{s})= Q_{k,l}\frac{\log r}{r^4}+O\left(\frac{1}{r^4}\right),
\end{equation}
where we find, after some algebra,
\begin{eqnarray}
\label{Lambda}
Q_{k,l} &=& \frac{(\pi - 2)^2}{4 \pi ^6}\Bigl(g_{k-1,l-1}-4 g_{k-1,l}+g_{k-1,l+1}\nonumber\\
&-&g_{k,l-2}+4 g_{k,l}-g_{k,l+2}-g_{k+1,l-1} \nonumber\\
&+&4 g_{k+1,l}-g_{k+1,l+1}-2 g_{k+2,l}\Bigr).
\end{eqnarray}
The summation over all $k$, $l$ yields
\begin{equation}
\label{ExtraTerms}
\sum_{k=-\infty}^{+\infty} \sum_{l=-\infty}^{+\infty}Q_{k,l} = \frac{(\pi-2)^2}{\pi^6}.
\end{equation}
Finally, we obtain
\begin{equation}
\label{AnswerTHeta}
 P_{1\Theta}(r) -P_1P_\Theta= -\frac{2(\pi-2)^2}{\pi^6}  \frac{\log r}{r^4}+O\left(\frac{1}{r^4}
\right),
\end{equation}
which coincides with the LCFT prediction (\ref{P1i}) for $i=2$. Similar calculations for
$P_{13}(r)$ and $P_{14}(r)$ fully confirms the results (\ref{P1i}) with the correct values
of the coefficients.

Despite a very specific form of $\Delta_{\Theta}$, the correlation functions $P_{1i}(r)$,
$i=2,3,4,$
are the first example where the logarithmic corrections to pair correlations
can be computed explicitly.

\section*{Acknowledgments}
This work was supported by a Russian RFBR grant, No 06-01-00191a, and by the Belgian
Internuniversity Attraction Poles Program P6/02. P.R. is a Research Associate of the
Belgian National Fund for Scientific Research (FNRS).

\end{document}